\RequirePackage{snapshot}
\PassOptionsToPackage{colorlinks=true,citecolor=blue,linkcolor=blue,urlcolor=airforceblue}{hyperref}

\documentclass[a4paper,superscriptaddress,11pt,unpublished,onecolumn]{quantumarticle}
\pdfoutput=1
\usepackage[utf8]{inputenc}
\usepackage[english]{babel}
\usepackage[T1]{fontenc}
\usepackage{amsmath}

\usepackage{xcolor}
\definecolor{lightblue}{rgb}{0.68, 0.85, 0.9}
\definecolor{airforceblue}{rgb}{0.36, 0.54, 0.66}

\usepackage{hyperref}

\usepackage{tikz}
\usepackage{lipsum}

\usepackage{caption}
\usepackage{accents}

\usepackage{subcaption}

\usepackage{parskip}

\usepackage{bbold}

\newcommand{\bite}{\begin{itemize}}
\newcommand{\eat}{\end{itemize}}
\newcommand{\beq}{\begin{equation}}
\newcommand{\eeq}{\end{equation}}

\newcommand{\beqa}{\begin{align}}
\newcommand{\eeqa}{\end{align}}
\newcommand{\barr}{\begin{array}}
\newcommand{\earr}{\end{array}}

\newcommand{\ie}{\textit{i.e.}~}
\newcommand{\eg}{\textit{e.g.}~}

\newcommand{\mb}[1]{\mathbf{#1}}
\newcommand{\mc}[1]{\mathcal{#1}}
\newcommand{\mbb}[1]{\mathbb{#1}}

\newcommand{\vect}[1]{\boldsymbol{#1}}
\newcommand{\expect}[1]{\left\langle #1 \right\rangle}

\newcommand{\ket}[1]{\vert #1 \rangle}

\newcommand\scalemath[2]{\scalebox{#1}{\mbox{\ensuremath{\displaystyle #2}}}}

\graphicspath{{figures/}}

\begin{document}

\title{Quantum Error Correction in Loop Quantum Gravity}
%\date{\today}
\date{Dec 26, 2019}
\author{Deepak Vaid}
\affiliation{National Institute of Technology Karnataka, Surathkal, India}
%\affiliation{Inter-University Centre for Astronomy and Astrophysics, Pune, India}
\email{dvaid7@gmail.com}
\homepage{https://www.quantumofgravity.com/blog}
%\orcid{0000-0002-0310-4599}
%\author{Sundance Bilson-Thompson}
%\affiliation{University of Adelaide, Adelaide, Australia}
%\email{sundancebt@gmail.com}
%\thanks{You can use the \texttt{\textbackslash{}email}, \texttt{\textbackslash{}homepage}, and \texttt{\textbackslash{}thanks} commands to add additional information for the preceding \texttt{\textbackslash{}author}. If applicable, this can also be used to indicate that a work has previously been published in conference proceedings.}
\maketitle

\begin{abstract}
Previous works (by Almiehri, Dong, Harlow, Pastakawski, Preskill, Yoshida and others) have established that quantum error correction plays an important role in understanding how the bulk degrees of freedom of an Anti-deSitter spacetime are encoded in the degrees of freedom of the boundary Conformal Field Theory. In previous work \cite{Vaid2013Elementary} I have argued that the Bilson-Thompson model \cite{Bilson-Thompson2006Quantum,Vaid2010Embedding} of elementary particles allows us to view elementary particles as gates for universal quantum computation. In the present work I show that the Bilson-Thompson model, where elementary particles are represented by elements of the framed braid group on three strands, provides an explicit model for the generation of qutrit (three-qubit) states which are the ingredients of Shor's quantum error correcting code. This allows, for the first time, to connect in a concrete manner the proposals of Almheiri, Pastawski, Preskill and others regarding the role of quantum error correction in quantum gravity, to a viable model of elementary particles. Loop Quantum Gravity (LQG), the theory of quantum gravity in which such topological excitations exist, can thus serve as the glue which can connect AdS/CFT based approaches to quantum gravity to the well understood physics of the Standard Model.
\end{abstract}

\tableofcontents

\section{Introduction}

In recent years, following the seminal works by several groups \cite{Almheiri2014Bulk,Mintun2015Bulk-Boundary,Pastawski2015Holographic,Freivogel2016Precursors} it has become clear that quantum error correction plays an important role in understanding how the bulk geometry of an AdS spacetime emerges from the conformal field theory living on the boundary of that spacetime.

\begin{figure}[h]
	\centering
	\begin{subfigure}[t]{0.4\textwidth}
		\centering
		\includegraphics[height=100pt]{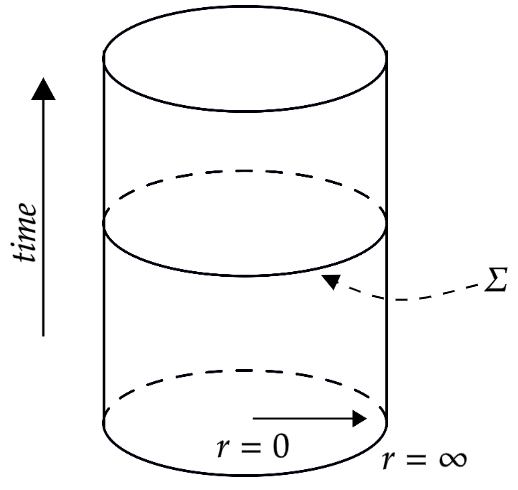}
		\caption{Conformal diagram of Anti-deSitter space. Time runs vertically upwards. In global co-ordinates $r=0$ represents the center of the spacetime and the boundary is located at $r=\infty$. $ \Sigma $ is a constant time slice.}
		\label{fig:ads-cylinder}
	\end{subfigure}
	\hfill
	\begin{subfigure}[t]{0.4\textwidth}
		\centering
		\includegraphics[height=100pt]{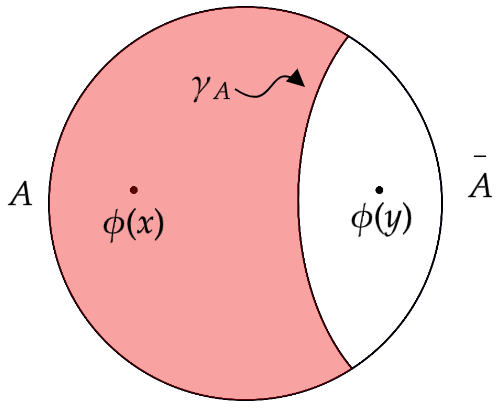}
		\caption{A single time-slice of the AdS cylinder. The boundary at $r=\infty$ is divided into two sections $A$ and $\bar{A}.$ $\gamma_{A}$ is the surface of minimum area anchored to the boundary and separating the two regions $A$ and $\bar{A}$.}
		\label{fig:ads-spatial-slice}
	\end{subfigure}
	\caption{Conformal diagram of AdS spacetime (left) and the Ryu-Takayanagi surface on a constant time-slice (right).}
	\label{fig:ads}
\end{figure}

\autoref{fig:ads} shows the conformal diagram of AdS space on the left and a constant time slice of this spacetime on the right. The surface $\gamma_A$ divides the bulk geometry into two regions, thus also dividing the boundary into two regions labeled $ A $ and $ \bar A $. According to the Ryu-Takayanagi conjecture \cite{Ryu2006Aspects,Ryu2006Holographic} the entropy due to entanglement between the degrees of freedom living in $ A $ with those living in $ \bar A$ is given by the area of the surface $ \gamma_A $ dividing the bulk interior:

\begin{equation}\label{key}
	S_{A,\bar A} = \frac{\text{Area of }\gamma_{A}}{4 G_N^{d+2}}
\end{equation}

where $ d $ is the number of spatial dimensions of the bulk AdS geometry.

\begin{figure}[h]
	\centering
	\begin{subfigure}[t]{0.4\textwidth}
		\centering
		\includegraphics[height=100pt]{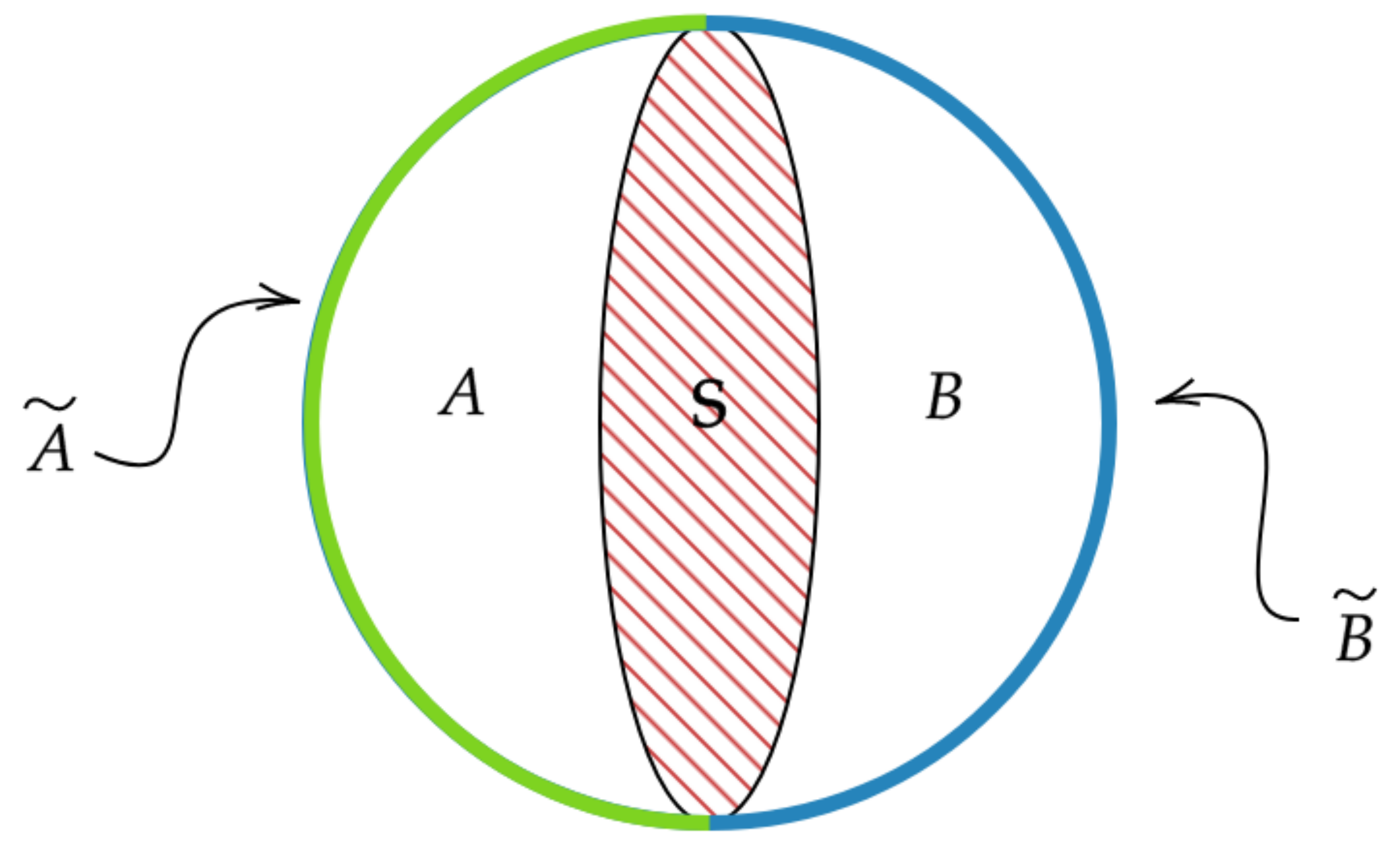}
		\caption{The surface $S$ divides the bulk geometry into two regions $A$ and $B$. The area of $S$ is a measure of the entropy of entanglement between the degrees of freedom living on the boundary of the regions $A$ and $B$ respectively.}
		\label{fig:divided-bulk}
	\end{subfigure}
	\hfill
	\begin{subfigure}[t]{0.4\textwidth}
		\centering
		\includegraphics[height=120pt]{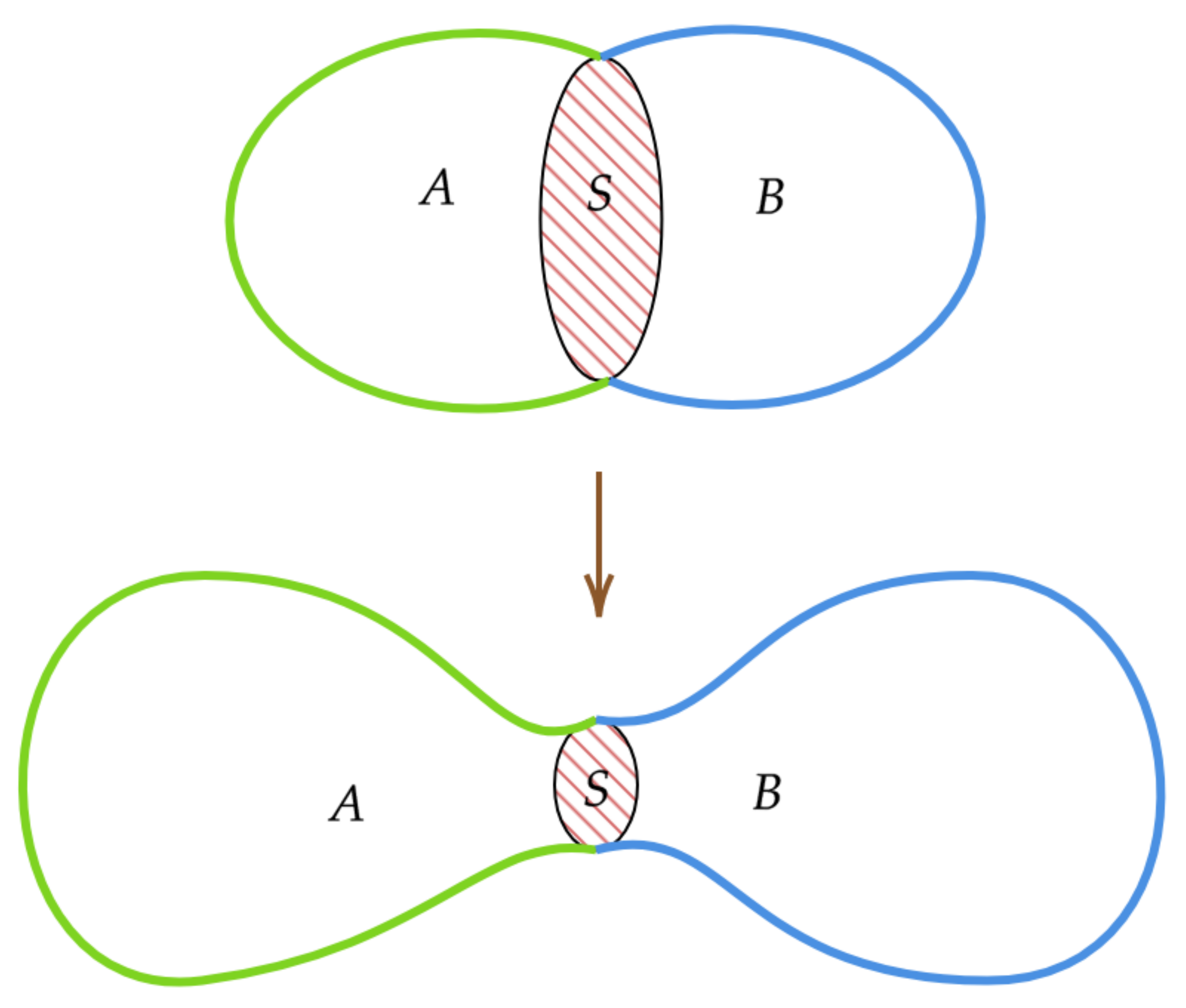}
		\caption{As we shrink the surface $S$, we reduce the entropy of entanglement between the two regions. Geometrically this corresponds to separating or cutting the bulk into disconnected regions.}
		\label{fig:shrinking-area}
	\end{subfigure}
	\caption{Illustrating the relation between geometric connectivity and entanglement \cite{Van-Raamsdonk2010Building}}
	\label{fig:ads-rt}
\end{figure}

The same time slice is shown in \autoref{fig:ads-rt}. The bulk is divided into two regions, now labeled $A$ and $B$, with the dividing surface labeled $S$. Now, imagine shrinking the surface $S$ so that its area gradually decreases. As the area of this surface reduces, the Ryu-Takayanagi formula tells us that the entropy of entanglement, between the degrees of freedom living on the boundary of region $ A $ and those living on the boundary of region $ B $, also starts to reduce. Geometrically, the effect of shrinking $ S $ is to reduce the connectivity between the bulk regions $A$ and $B$ as shown on the right side of \autoref{fig:ads-rt}. When the area of $S$ goes to zero, the two regions will become completely disconnected. At the same time the entropy of entanglement between the degrees of freedom living on the boundary of region $A$ with those living on the boundary of region $B$ will also vanish.

This simple thought experiment was first proposed in 2010 by Mark Van Raamsdonk \cite{Van-Raamsdonk2010Building}. It illustrates in a very simple yet dramatic manner how the RT formula, in conjunction with the AdS/CFT conjecture provides a straightforward description of how classical spacetime can emerge from entanglement between degrees of freedom of some underlying quantum field theory. It provides a very simple, yet physically plausible explanation for how different pieces of a spacetime can be ``sewn'' together to build a macroscopic geometry.

\subsection{AdS/CFT Bulk Reconstruction}

Further background for the problem requires us to use slightly more technical language. The metric for asymptotically Anti-deSitter spacetimes (AdS) is generally expressed in terms of the time-like co-ordinate $ t $, a ``radial'' co-ordinate $ r $ and a set of co-ordinates $ \vec{x} $ which describe the spatial dimensions orthogonal or transverse to $ r $: $ g(r,\vec{x},t) $. Using the global co-ordinates, the metric can be expressed as:

\begin{equation}\label{eqn:ads-global-metric}
	ds^2 = -\left( 1 + \frac{r^2}{L^2} \right) dt^2 + \left( 1 + \frac{r^2}{L^2} \right)^{-1} dr^2 + r^2 d\vec{x}^2
\end{equation}
where $ L $ is the AdS radius. This metric is a solution to the vacuum Einstein equations with a negative cosmological constant $ \Lambda < 0 $, given by: $ \Lambda = -d(d-1)/2L^2 $, where $ d $ is the total number of spacetime dimensions. In the AdS/CFT correspondence one usually talks about spacetimes which are only \emph{asymptotically} AdS, \ie whose metric can be written in the form:
\begin{equation}\label{eqn:asymptotic-ads-global-metric}
	ds^2 = -f(r) dt^2 + f(r)^{-1} dr^2 + r^2 d\vec{x}^2
\end{equation}
where $ f(r) = 1 + r^2/L^2 $ for $ r $ large enough. The boundary of the spacetime is located at $ r = \infty $.

Consider some asymptotically AdS spacetime $ \mc{M} $, with boundary $ \partial \cal M $. The bulk geometry of $ \cal M $ satisfies the Einstein field equations. Consider some fields $ \phi(r,\vec{x}) $ living on a constant time slice $ \Sigma_t $ of $ \cal M $. In terms of these quantities the AdS/CFT conjecture can be stated mathematically in terms of the following expression, known as the Gubser-Klebanov-Polyakov-Witten (GKP-Witten) relation \cite{Gubser1998Gauge,Witten1998Anti}:

\begin{equation}\label{eqn:ads-cft-identity}
	\lim\limits_{r \rightarrow \infty} e^{\imath S[\phi(r,\vec{x})]} \equiv \expect{\exp \left( \imath \int \phi_0(\vec{x}) \mc{O} \right)}
\end{equation}
where $ S[\phi(r,\vec{x})] $ is the gravitational action evaluated in the bulk for a given field configuration $ \phi(r,\vec{x}) $; $ \phi_0(\vec{x}) $ is the value of the fields evaluated only on the boundary and $ \mc{O} $ is some operator acting on the on the Hilbert space of the boundary field theory. This relation can be understood as the statement that:
\begin{quote}
	\textbf{Proposition I}: \emph{classical expectation values of bulk fields in asymptotically anti-deSitter spacetimes can be calculated in terms of quantum expectation values of operators acting on the Hilbert space of the boundary field theory}
\end{quote}

There is, however, a caveat which prevents us from constructing a one-to-one correspondence of the physics in the boundary with the physics in the bulk. This is because the equality \eqref{eqn:ads-cft-identity} holds only \emph{asymptotically} as $ r \rightarrow \infty $. The question that therefore presents itself is: \emph{can we reconstruct bulk fields (for $ r < \infty $) from knowledge of boundary fields?}

\begin{figure*}
	\centering
	\begin{subfigure}[t]{0.3\textwidth}
		\centering
		\includegraphics[height=100pt]{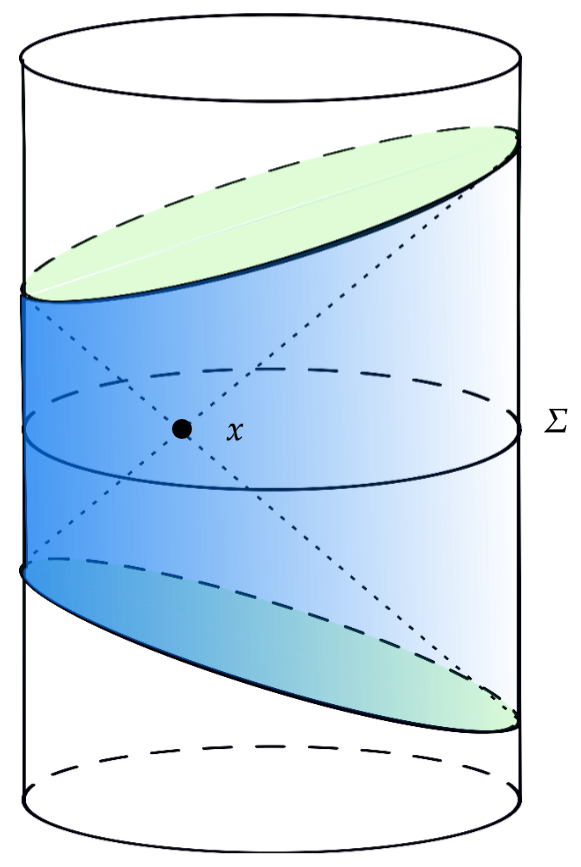}
		\caption{Local field defined at the bulk point $ x $, can be expressed \cite{Hamilton2006Holographic} in terms of non-local, boundary operators with support on $ \Sigma $ - the set of points on the boundary which are at spacelike separations from $ x $}
		\label{fig:bulk-boundary-p1}
	\end{subfigure}
	\hfill
	\begin{subfigure}[t]{0.3\textwidth}
		\centering
		\includegraphics[height=100pt]{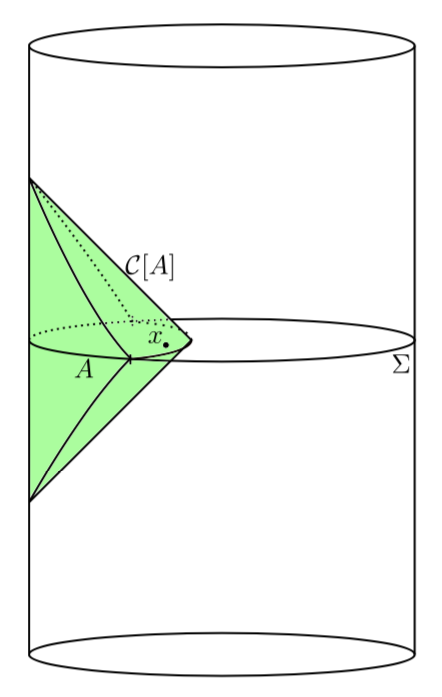}
		\caption{AdS-Rindler wedge reconstruction \cite{Bousso2012Light-sheets,Czech2012The-gravity,Hubeny2012Causal}. Requires knowledge of operators only in subset $ A \subset \Sigma $ of boundary Cauchy surface.}
		\label{fig:bulk-boundary-p2}
	\end{subfigure}
	\hfill
	\begin{subfigure}[t]{0.3\textwidth}
		\centering
		\includegraphics[height=100pt]{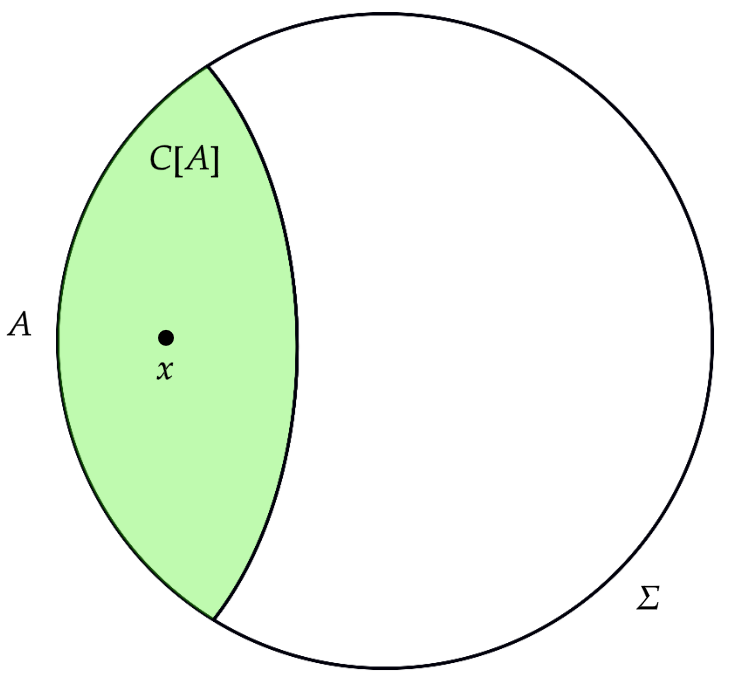}
		\caption{Fields at any point $ x $ in the green region can be reconstructed using operators living on $ A $. Ambiguity arises because $ x $ can lie in more than \emph{one} causal wedge.}
		\label{fig:bulk-boundary-p3}
	\end{subfigure}
	\caption{Reconstruction of bulk fields from boundary CFT operators}
	%        \label{fig:three graphs}
\end{figure*}

It turns out that bulk fields, \ie $ \phi(t,r,\vec{x}) $ with $ r < \infty $, can indeed be reconstructed from boundary operators \cite{Hamilton2006Holographic}. For this purpose it is helpful to work in the Poincare co-ordinates for AdS:
\begin{equation}\label{eqn:ads-poincare}
	ds^2 = \frac{1}{z^2}(-dt^2 + dz^2 + d\vec{x}^2)
\end{equation}
where now the radial co-ordinate is $ z $ and the boundary lies at $ z = 0 $. In these coordinates, on a constant time-slice, a bulk field with normalizable fall-off near the boundary can be written as:
\begin{equation}\label{eqn:field-normalizable-falloff}
	\phi(z, \vec{x}) \sim z^{\Delta} \phi_0(\vec{x})
\end{equation}
In terms of these co-ordinates, the value of the field at a point $ (z,\vec{x}) $ in the bulk can be expressed in terms of the boundary field via:
\begin{equation}\label{eqn:bulk-fields}
	\phi(z, \vec{x}) = \int_{\partial M} dx' K(\vec{x}'|z, \vec{x}) \phi_0(\vec{x}')
\end{equation}
where $ K $ is the kernel or the smearing function. $\phi_0(\vec{x})$ corresponds to a local operator $ \mc{O}(\vec{x}) $ in the CFT:
\begin{equation}\label{eqn:field-op-p1}
	\phi_0(\vec{x}) \leftrightarrow \mc{O}(\vec{x})
\end{equation}
This relationship implies that \emph{local} bulk fields are dual to \emph{non-local} boundary operators:
\begin{equation}\label{eqn:field-op-p2}
	\phi(z,\vec{x}) \leftrightarrow \int dx' K (\vec{x}'|z, \vec{x}) \mc{O}(\vec{x}')
\end{equation}
where the integral has support over a subset of points $ \vec{x}' $ on the boundaryConsider a spatial surface $ \Sigma $ which is a constant time slice of the AdS cylinder as shown in \autoref{fig:bulk-boundary-p1}. Then the integral in \eqref{eqn:field-op-p2} has support on the subset of the boundary shaded in green, \ie all the points on the boundary which are at a spacelike separation from the bulk point $ (z,\vec{x}) $. This construction can be further restricted \cite{Bousso2012Light-sheets,Czech2012The-gravity,Hubeny2012Causal} to limit the domain of integration to the portion of the boundary which lies in the ``causal wedge'' of $ (x,\vec{x}) $. This is the boundary region shaded in green shown in \autoref{fig:bulk-boundary-p2}. Here $ A \subset \Sigma $ is a subset of the spatial slice $ \Sigma $, such that the bulk point lies in the region enclosed by $ A $ and the Ryu-Takayanagi surface corresponding to $ A $ as shown in \autoref{fig:bulk-boundary-p3}.

\subsection{Redundancy in Bulk Reconstruction}

\begin{figure*}
	\centering
	\begin{subfigure}[t]{0.3\textwidth}
	\centering
	\includegraphics[height=100pt]{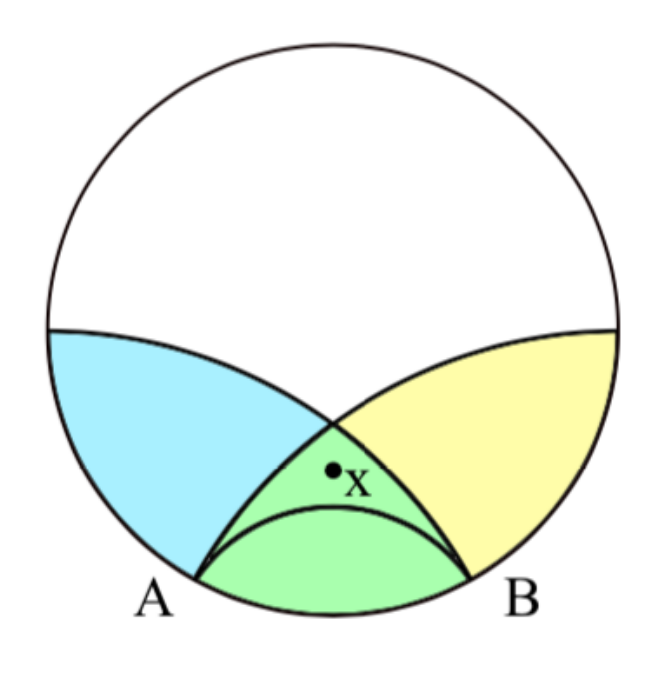}
	\caption{Fields at a point $ x $ can be represented in terms of operators which lie \emph{either} on the segment of the boundary labeled $ A $ \emph{or} on the segment labeled $ B $.}
	\label{fig:overlapping-wedges}
	\end{subfigure}
	\hfill
	\begin{subfigure}[t]{0.3\textwidth}
		\centering
		\includegraphics[height=100pt]{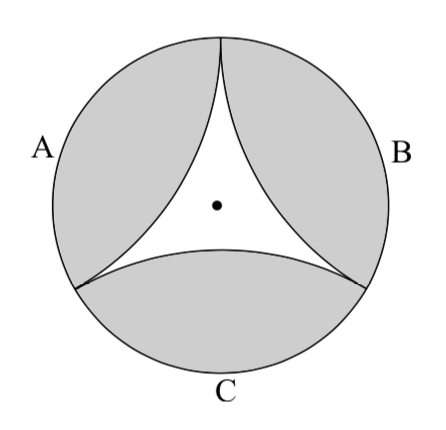}
		\caption{Three boundary regions $ A,B,C $ such that the bulk point does not lie in the causal wedge of any of the three regions (shown in grey).}
		\label{fig:non-overlapping-wedges}
	\end{subfigure}
	\hfill
	\begin{subfigure}[t]{0.3\textwidth}
		\centering
		\includegraphics[height=100pt]{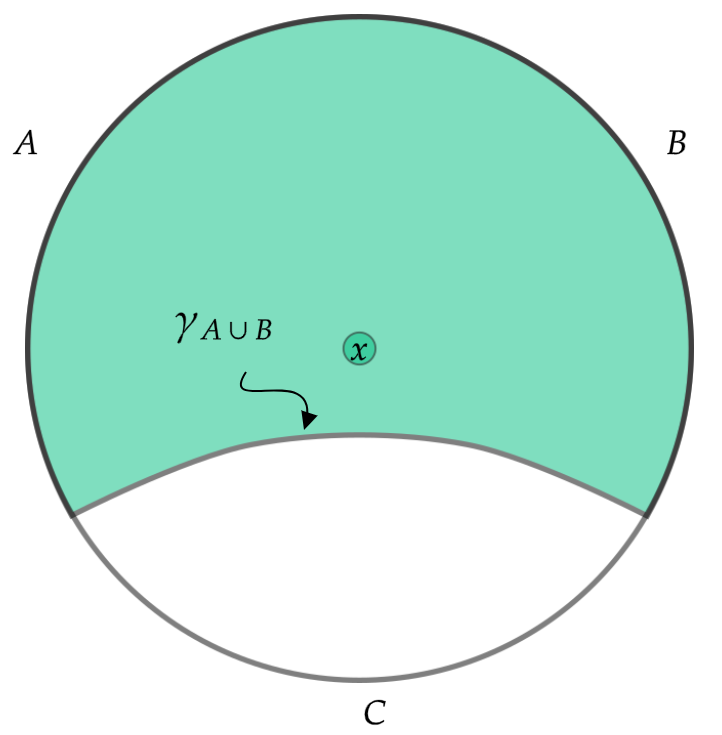}
		\caption{The bulk point \emph{does} lie within the causal wedges of regions $ A \cup B $, $ B \cup C $ and $ C \cup A $ separately. Here the causal wedge of $ A \cup B $ is shown shaded in green, along with the associated RT surface $ \gamma_{A \cup B} $.}
		\label{fig:wedge-union}
	\end{subfigure}
	\caption{Illustrating the ambiguity in the contruction of bulk fields in terms boundary operators in AdS/CFT}
	%        \label{fig:three graphs}
\end{figure*}

It now becomes apparent that there is a redundancy in this description of bulk reconstruction. Consider for instance a second region $ B $ lying on the boundary of $ \Sigma $, which has non-zero overlap with $ A $ as shown in \autoref{fig:overlapping-wedges}, such that the bulk point $ (z,\vec{x}) $ lies within the causal wedges of both $ A $ \emph{and} $ B $. According to the HRT prescription for bulk reconstruction, fields at the bulk point $ (z,\vec{x}) $ can be mapped \emph{either} to an operator $ \mc{O}_A[\phi(x)] $ with support \emph{only} in $ A $, \emph{or} to an operator $ \mc{O}_B[\phi(x)] $ with support \emph{only} in the region $ B $.

%\begin{figure}
%	\centering
%	\includegraphics[scale=0.6]{ads-rindler-p2}
%	\caption{Fields at a point $ x $ can be represented in terms of operators which lie \emph{either} on the segment of the boundary labeled $ A $ \emph{or} on the segment labeled $ B $.}
%	\label{fig:overlapping-wedges}
%\end{figure}

Now consider the boundary segment which consists of the intersection $ A \cap B $. The associated causal wedge $ W_C[A \cap B] $ \emph{does not} contain the point $ (z,\vec{x}) $, even though as is clear, the bulk point does fall within the intersection $ W_C[A] \cup W_C[B] $ of the causal wedges of $ A $ and $ B $. In order for there to be a \emph{unique} description of the bulk fields in terms of boundary operators we require that:
\begin{equation}\label{eqn:equiv-ops}
\mc{O}_A[\phi(x)] = \mc{O}_B[\phi(x)]
\end{equation}
For this to be true, both operators must have support \emph{only} in $ A \cap B $. However, as just noted the causal wedge of this region $ W_C[A \cap B] $ does not contain the bulk point and therefore any CFT operators with support \emph{only} on $ A \cap B $ cannot provide a faithful representation of bulk fields at $ (z,\vec{x}) $. The only conclusion one can arrive at is that:
\begin{equation}\label{eqn:inquiv-ops}
	\mc{O}_A[\phi(x)] \ne \mc{O}_B[\phi(x)]
\end{equation}
Since both operators \emph{separately} encode the physics near the same bulk point and the fact that the two operators \emph{cannot} be the same, leads us to conclude that \cite{Almheiri2014Bulk}:
\begin{quote}
	\textbf{Proposition II}: \emph{There does not exist any \textbf{unique} representation of bulk fields in terms of CFT operators living on (some subset of) the boundary.}
\end{quote}

We can shed more light on the nature of this redundancy by considering three segments of the boundary as shown in \autoref{fig:non-overlapping-wedges}. As can be seen the bulk point (in the center) \emph{does not} lie within the causal wedges (shaded in grey) of either of the three regions $ A,B $ or $ C $. Therefore we do not expect to have a representation of bulk fields at the given point to have a representation in terms of operators defined solely on either one of the three regions. Now consider, the third figure \autoref{fig:wedge-union}. From this we can see that the bulk point now lies in the causal wedge of the regions $ A \cup B $, $ B \cup C $ and $ C \cup A $, taken \emph{separately}. Consequently there exists a representation of the bulk fields in terms of operators: 
\[ \mc{O}_{A \cup B} \ne \mc{O}_{B \cup C} \ne \mc{O}_{C \cup A} \]
As Almiehri et. al. point out in \cite{Almheiri2014Bulk} this sort of dependence of the bulk fields on three different operators defined on overlapping boundary regions is reminiscent of the three-qutrit quantum error correcting code wherein one ``logical'' qutrit is encoded in terms of three ``physical'' qutrits. At this stage, let us rapidly overview the basic ideas behind quantum error correcting codes and the three-qutrit code in particular.
	            
\section{Quantum Error Correction}

In any computation, quantum or classical, errors are inevitable. ``Error correction'' protocols \cite{Kempe2006Approaches,Gottesman2009Introduction} allow us to correct (non-catastrophic) errors. The basic idea behind all error correction protocols, whether classical or quantum, is to encode the ``logical'' degrees of freedom - which carry the information we would like to protect from errors - in terms of some ``physical'' degrees of freedom. The number of physical d.o.f is necessarily larger than the number of logical d.o.f. The simplest example of such a scheme is the repetition code, for either classical or quantum information, in which a single physical d.o.f is encoded in two or more logical d.o.f in the following manner:
\begin{align}
	\text{Classical}: ~ & \tilde 0 \rightarrow 000; \quad \tilde 1 \rightarrow 111 \nonumber \\
	\text{Quantum}: ~ & \ket{\tilde 0} \rightarrow \ket{000}; \quad \ket{\tilde 1} \rightarrow \ket{111}
\end{align}
where symbols with a $ \tilde{~} $ on top represent logical qubits and those without it are physical qubits. Here we have encoded one logical cbit/qubit in three physical cbits/qubits. Let us assume that the errors in our system will only flip one of the physical cbits/qubits in a single code-word, e.g. $ 000 \rightarrow 010 $. Such errors can therefore be corrected by using the \emph{majority rule}. Those cbits/qubits which are greater in number are the ``correct'' ones and those which are in the minority are the ``wrong'' ones. Thus $ 010 $ would become $ 000 $ after correction and $ 101 $ would become $ 111 $. Of course, if the errors cause more than one cbit/qubit to flip then this simple scheme would no longer be effective.

The basic principle behind all forms of error correction is \emph{safety through redundancy}.

\textbf{Definition:} A $ (N,K) $ quantum error correcting code consists of a triple \cite{Kribs2005Unified}:
\begin{equation}\label{eqn:qec-triple}
	(\mc{C}_K, \mc{E}, \mc{R})
\end{equation}
where: $ \mc{C}_K $ is the ``code space'', $ K $ dimensional subspace of a higher dimensional ``physical'' Hilbert space $ \mc{C}_K \subset \mc{H}_N $; $ \mc{E} $ is a set of quantum operations on $ \mc{H}_N $ which generate errors; and $ \mc{R} $ is a set of quantum operations on $ \mc{H}_N $ which can ``undo'' the errors. $ N, K $ denote, respectively, the dimensions of the physical Hilbert space and the code subspace.

An example of such a code is Shor's nine qubit code, where a single logical qubit is encoded using nine physical qubits.

\begin{align}\label{eqn:nine-qubit-code}
	\ket{\tilde 0} & = \frac{1}{2\sqrt{2}} \left( \ket{000} + \ket{111} \right)\left( \ket{000} + \ket{111} \right)\left( \ket{000} + \ket{111} \right) \nonumber \\
	\ket{\tilde 1} & = \frac{1}{2\sqrt{2}} \left( \ket{000} - \ket{111} \right)\left( \ket{000} - \ket{111} \right)\left( \ket{000} - \ket{111} \right)
\end{align}

The fundamental ingredient in this code are the GHZ (Greenberger-Horne-Zeilinger) states\footnote{These states are also sometimes referred to as ``cat'' states in honor of Schrodinger's famous feline}:
\begin{equation}\label{eqn:ghz-states}
	\ket{\pm} = \frac{1}{\sqrt{2}}(\ket{000} \pm \ket{111})
\end{equation}

For future reference we reproduce below the quantum circuit which is used for generating GHZ states:

\begin{figure}[h]
	\centering
	\includegraphics[scale=1.0]{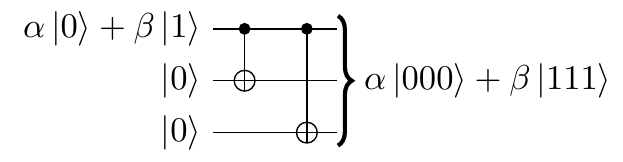}
	\caption{Quantum circuit for generating a qutrit state using CNOT gates}
	\label{fig:GHZ-states}
\end{figure}

This circuit involves application of the CNOT gate to qubits $ (1,2) $ and qubits $ (1,3) $ in succession. The CNOT gate is shown below:

\begin{figure}[h]
	\centering
	\includegraphics[origin=c]{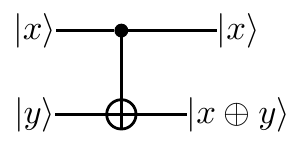}
	\caption{CNOT gate. Here $ x $ and $ y $ take values in $ \{0,1\} $ and $ \oplus $ is the logical XOR operator. If the control qubit $ \ket{x} = \ket{0} $ then the target qubit $ \ket{y} $ is left unchanged, otherwise the target qubit is flipped.}
	\label{fig:cnot-gate}
\end{figure}

In the remainder of this paper we will show that the consideration of discrete symmetries of	 spin-networks - the kinematical states of quantum geometry in the framework of Loop Quantum Gravity - naturally leads us to discover the existence of topological excitations which correspond precisely to CNOT gates. These excitations can then be used to build GHZ states, or in combination with certain other single qubit gates - also represented in terms of operators acting on topological degrees of freedom - can be be used to generate an arbitrary quantum circuit.

\section{Spin Networks and  Quantum Geometry}

\begin{figure}[h]
	\centering
	\begin{subfigure}[t]{0.45\textwidth}
		\centering
		\includegraphics[height=100pt]{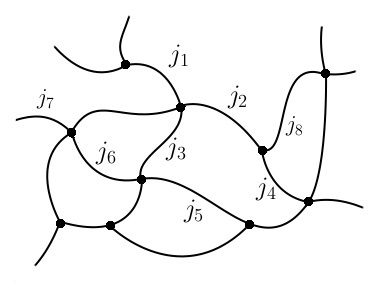}
		\caption{}
		\label{fig:spin-network}
	\end{subfigure}
	\hfill
	\begin{subfigure}[t]{0.45\textwidth}
		\centering
		\includegraphics[height=100pt]{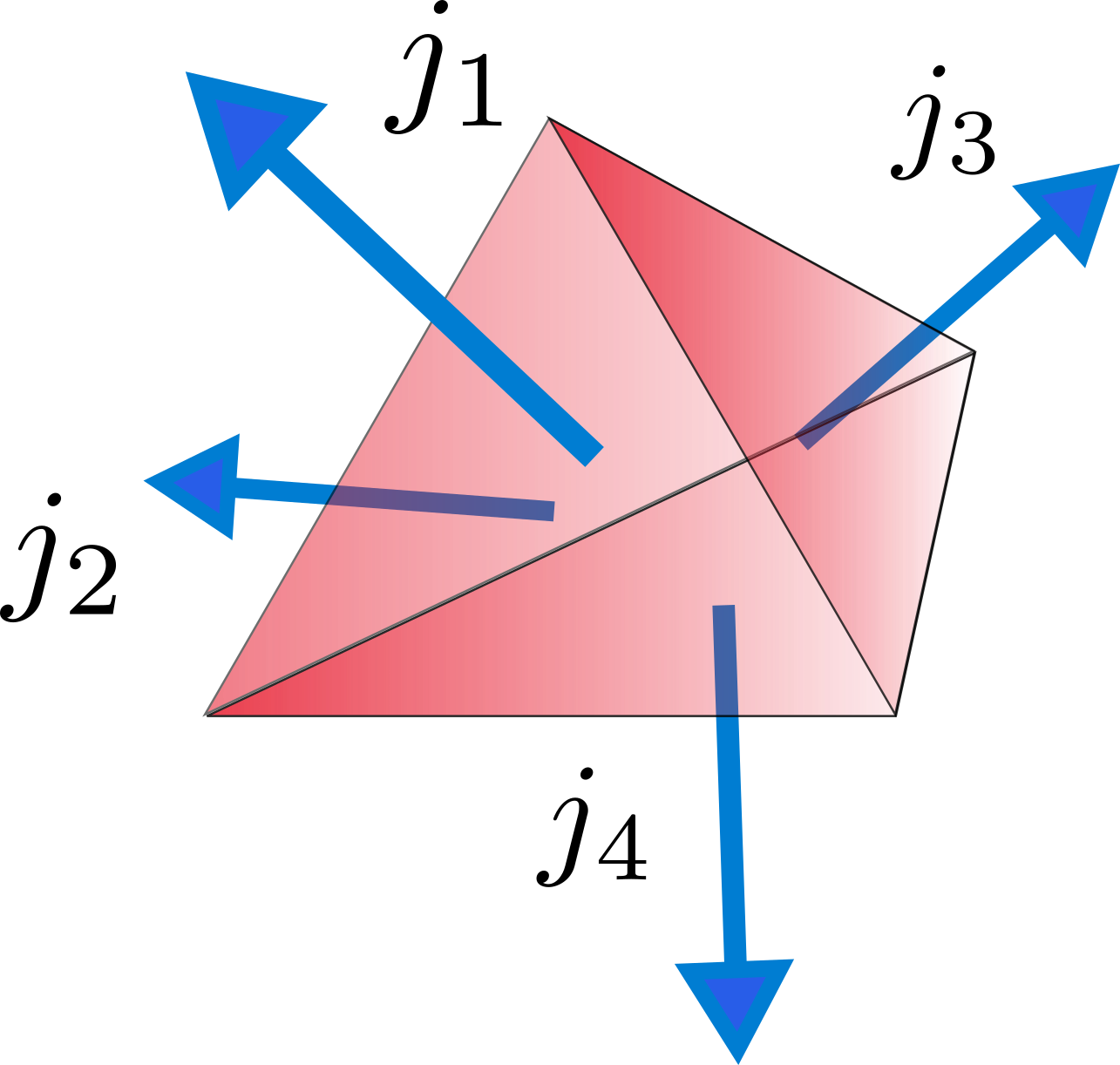}
		\caption{}
		\label{fig:spin-network-node}
	\end{subfigure}
	\caption{Shown on the left is a spin-network and on the right is a single (tetravalent) vertex.}
\end{figure}

Loop Quantum Gravity - or ``LQG'' for short - is an approach\footnote{there are many introductory reviews on LQG, at various levels of technical sophistication ranging from advanced \cite{Thiemann2001Introduction,Thiemann2002Lectures, Ashtekar2004Background} to intermediate \cite{Mercuri2010Introduction,Dona2010Introductory,Esposito2011An-introduction,Ashtekar2012Introduction} to (relatively speaking) elementary \cite{Vaid2014LQG-for-the-Bewildered,Rovelli2011Zakopane,Rovelli2014Covariant}} towards building a non-perturbative theory of quantum gravity. LQG provides a description of quantum states of geometry in terms of objects known as ``spin networks''. A spin-network is an arbitrary graph $ \Gamma $, whose edges are labeled by representations of $ SU(2) $ (\autoref{fig:spin-network}) and whose vertices are labeled by invariant $ SU(2) $ tensors known as ``intertwiners''. More simply, spin-networks are graphs whose edges carry angular momenta and whose vertices provide a means for adding together all the angular momenta coming into that vertex from its adjoining edges.

\begin{figure}[h]
	\centering
	\begin{subfigure}[t]{0.45\textwidth}
		\centering
		\includegraphics[height=100pt]{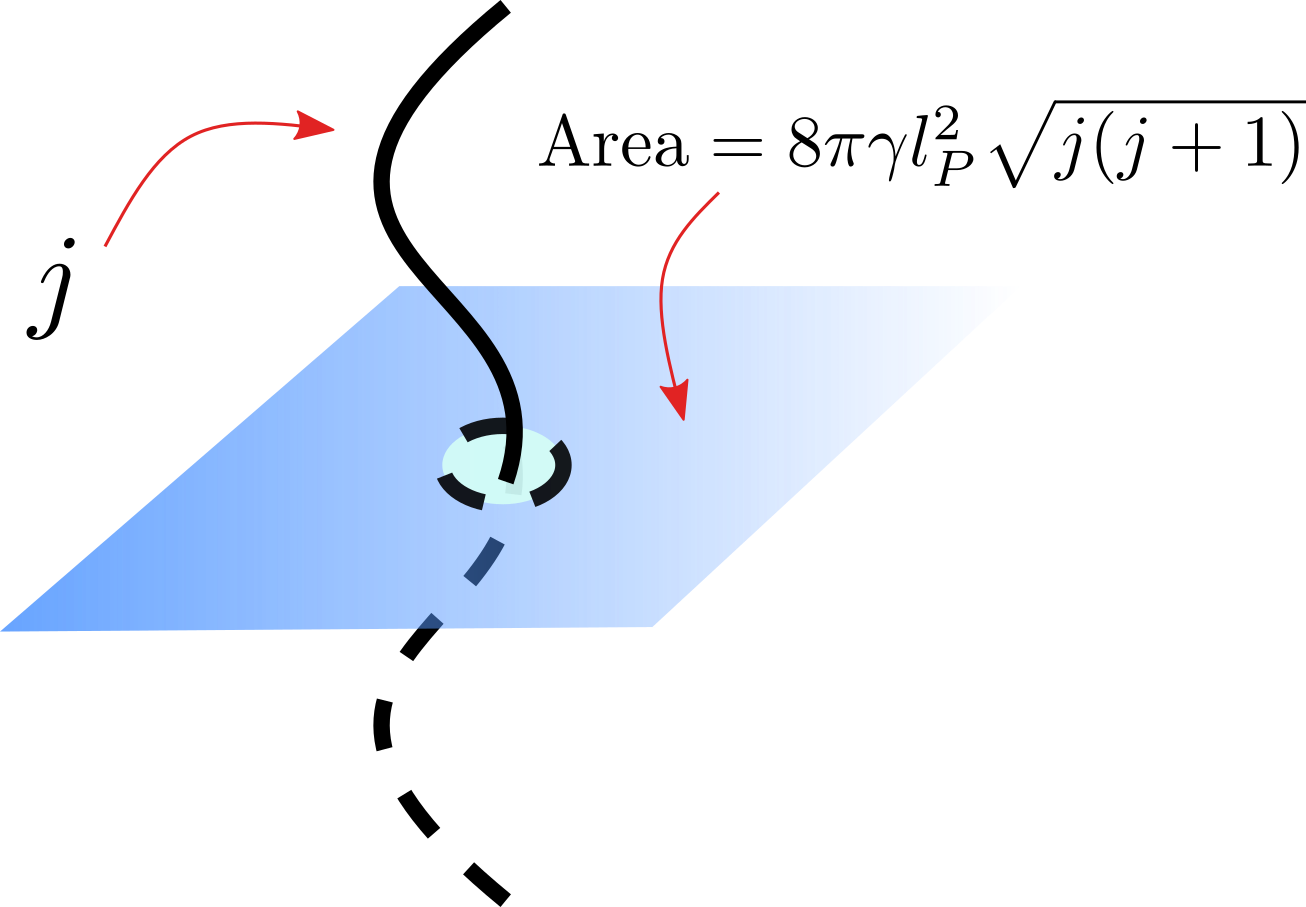}
		\caption{}
		\label{fig:area-puncture}
	\end{subfigure}
	\hfill
	\begin{subfigure}[t]{0.45\textwidth}
		\centering
		\includegraphics[height=100pt]{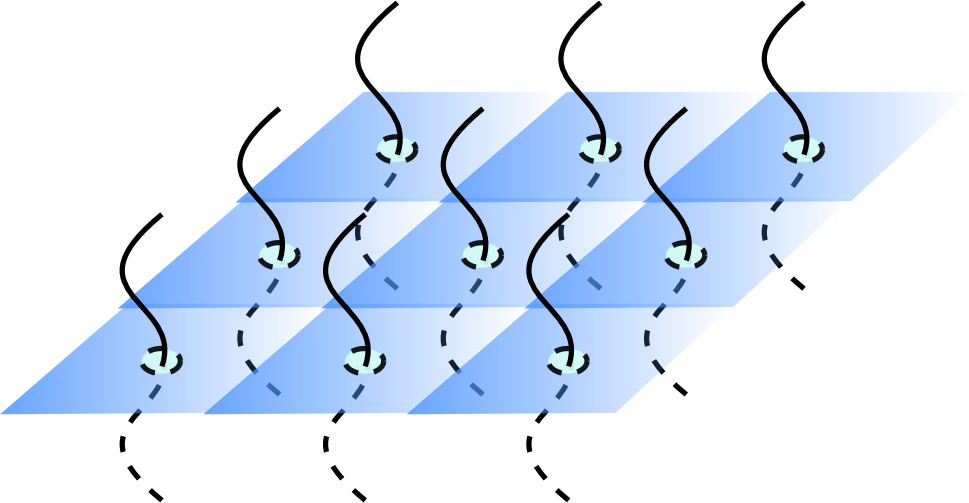}
		\caption{}
		\label{fig:area-punctures}
	\end{subfigure}
	\caption{Shown on the left is the action of the area operator and on the right an illustration of how a macroscopic surface can be built up from ``gluing'' together many quanta of area.}
\end{figure}

A spin-network defines a state of quantum geometry. With each edge we can associate a quantum of area and with each vertex a quantum of volume. \autoref{fig:area-puncture} shows a single edge of a spin-network carrying an angular momentum $ j $, puncturing a surface (shaded in blue). This edge endows the surface with a quantum of area given by $ A_j = 8\pi \gamma l_{P}^2 \sqrt{j(j+1)} $, where $ l_{P} $ is the Planck length and $ \gamma $ is a free parameter known as the Barbero-Immirzi parameter. The particular value of $ \gamma $ will not be relevant for our purpose.

Spin network vertices are associated with quanta of volume. In \autoref{fig:spin-network-node} a single tetravalent vertex is ``blown up''. Each edge adjoining the vertex carries an angular momentum $ j_i $. With each edge we can associate a triangle of area $ A_{j_i} $. The four triangles associated with the given vertex, will then close up\footnote{The four angular momenta must satisfy the closure condition $ \sum_{i=1}^4 \vect{j_i} = 0  $. This is analogous to the requirement that a classical tetrahedron satisfies $ \sum_{i=1}^4 \vect{N_i} = 0 $, where $ \vect{N_i} $ are the vectors normal to each face.} to form a tetrahedron with which we can associate an operator whose eigenvalues can be interpreted as the volume of the resulting ``quantum tetrahedron''. For more details we direct the interested reader to one of the several reviews listed in the references.

Spin networks provide a \emph{genuinely non-perturbative} way to quantize classical geometry. \emph{A priori}, there is no requirement for a background manifold with any sort of topological or geometric structure, in order to be able to define a quantum state of geometry. The only information required to specify a spin-network state $ \ket{\Psi_\Gamma} $ is knowledge of the graph $ \Gamma $ (\ie the number of edges and vertices and their connectivity structure) and the labeling of edges and vertices by spins and intertwiners respectively:

\begin{equation}\label{eqn:spin-net-info}
	\ket{\Psi_\Gamma} \equiv (\Gamma, N_v, N_e, A_{pq}, \{j_i\}, \{I_p\})
\end{equation}
where $ N_v, N_e $ are the numbers of vertices and edges respectively in the graph, $ A_{pq} $ is the connectivity matrix, $ \{j_i\} $ are the edge labels and $ \{ I_p \} $ are the vertex labels.

Now what makes these states \emph{bona-fide} states of quantum geometry, rather than simply being an \emph{ad-hoc} construction is the fact that these states are \emph{exact} solutions of the ADM Hamiltonian. As is very well understood, in the $ 3+1 $ formulation of general relativity, first presented by Arnowitt, Deser and Misner \cite{Arnowitt2004The-Dynamics} Einstein's equations can be expressed as a sum of constraints. When working in the connection formulation there are three constraints. These are known as the Gauss constraint $ \mc{C} $, the vector or ``diffeomorphism'' constraint $ \mc{H}_{diff} $ and the scalar or ``Hamiltonian'' constraint $ \mc{H}_{scalar} $. In this language the problem of quantum gravity reduces to that of finding states which are annihilated by the quantum operator versions of all three constraints. It turns out that spin network states are annihilated by two of the constraints - the Gauss and diffeomorphism constraint. The Gauss constraint is the statement of invariance of the state under $ SU(2) $ gauge transformations. This is satisfied by spin network states due to the closure condition: $ \sum j_e = 0 $, \ie the sum of angular momenta carried by all the edges adjoining a given vertex must give zero. The diffeomorphism constraint is trivially satisfied because the embedding of a graph in a smooth manifold does not affect its geometric content. One can move the edges and vertices of the graph around in the ambient manifold, but as long as the connectivity structure of the graph is unchanged, the spin network state is also unaffected.

The action of the third constraint - the Hamiltonian constraint - is expected to generate ``time evolution'' of the quantum geometry. The resulting states are known as ``spin foams'' and represent causal histories connecting two different spin network configurations. For our purposes we are interested only in spin-networks which represent the spatial geometry as a given instance of time.

\section{Diffeomorphism Invariance of Spin Networks}

The ADM constraints discussed in the last section fail to capture the full dynamics of general relativity. The reason for this is the existence of ``large'' diffeomorphisms (those which cannot be continuously deformed to the identity). This is completely analogous to the existence of topologically non-trivial solutions in field theory due to the existence of ``large'' gauge transformations. Diffeomorphisms constitute the gauge symmetry of general relativity. Therefore it is natural to consider not only small diffeomorphisms - those continously deformable to the identity - but also large diffeomorphisms - those which lie in the disconnected component of the gauge group. The ADM constraints only encode the small diffeomorphisms. As was discovered by Freidman and Sorkin in the 1980s \cite{Friedman1982Half-integral}, the consideration of large diffeomorphisms leads to manifolds which have half-integer spin.

In order to complete the description of spin network states as candidate states of quantum geometry, one has to consider the action of large diffeomorphisms in addition the small ones. Spin networks are trivially invariant under small (spatial) diffeomorphisms, because as explained in the previous section, moving edges or vertices around without changing the graph connectivity or spin labels does not change the geometric content of the state. Now, large diffeomorphisms would also \emph{not} change the graph connectivity or the spin labels. Thus the questions arises as to what effect, if any, would such transformations have on spin networks? It is not hard to convince oneself that the effect of large diffeomorphisms would be to break and reconnect spin-network edges in such a way that the winding and linking numbers of those edges would change. Thus, while the graph structure and spin labels would remain intact, a spin-network with two of its edges ``braided'' around one another is a topologically distinct state from one which does not have such a braiding.

\begin{figure}[h]
	\centering
	\includegraphics[scale=0.20]{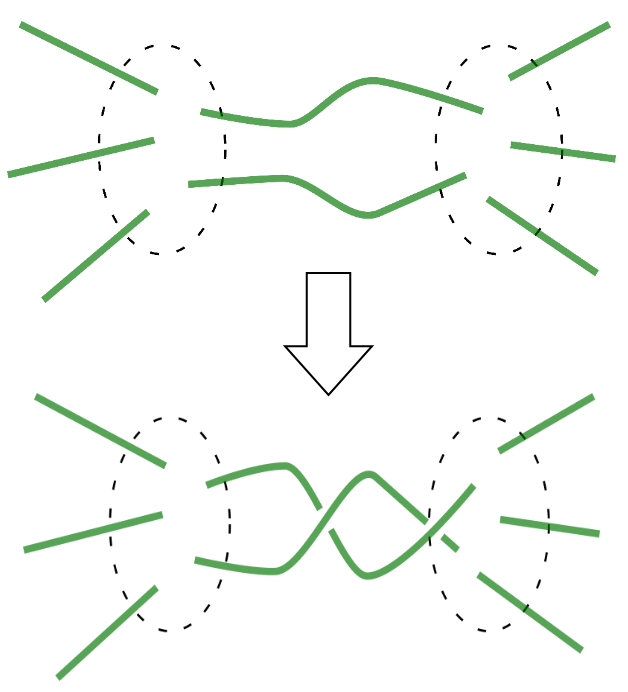}
	\caption{Action of a large diffeomorphism on a spin network state}\label{fig:large-diffeos}
\end{figure}

This is illustrated in \autoref{fig:large-diffeos}. The upper portion of the figure shows a generic spin network state with two edges connecting two different regions (whose boundary is drawn as the dotted curve) of the graph. The lower portion shows what happens to this graph under the action of a large diffeomorphism. The two edges break and reconnect after having wound around each other once. The resulting graph carries the same geometric information as the initial state but has a different topological structure.

If we wish to construct states of quantum geometry which incorporate the \emph{full} dynamics of quantum general relativity, we must construct states which are invariant under \emph{both} small \emph{and} large diffeomorphisms.

\section{Diffeomorphism Invariance and Yang-Baxter Equation}

\begin{figure}[h]
	\centering
	\includegraphics[height=100pt]{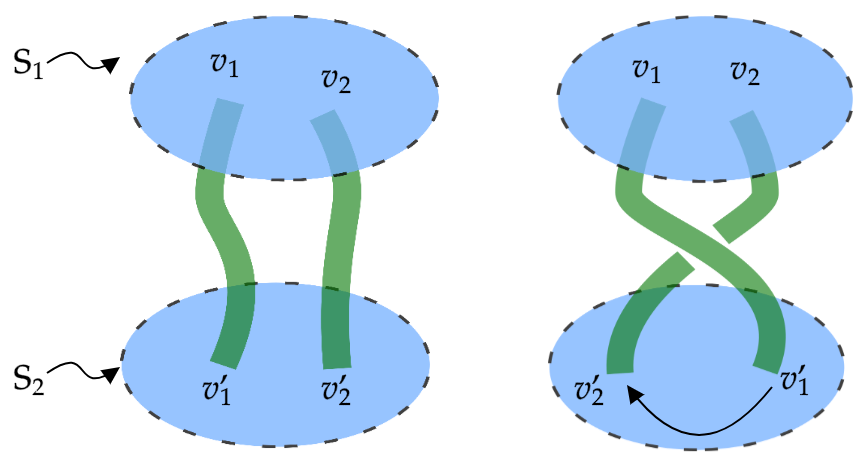}
	\caption{Two configurations of a pair of edges of some graph state $ \ket{\Psi_\Gamma} $. The left side shows the edges as being unbraided. On the right the same two edges are braided once around each other.}
	\label{fig:braiding}
\end{figure}

Now, in order to understand the effect of large diffeomorphisms on a quantum state of geometry, let us consider the two configurations depicted in figure \autoref{fig:braiding}. There are two surfaces $ S_1 $ and $ S_2 $ with two edges (shown in green) connecting them. Now, as shown for \eg in \cite{Ghosh2014CFT/Gravity}, we know that at the points where the edges punctures the surfaces we can associate the Hilbert space of a point particle. Let us denote the state of the two punctures on the first surface as $ \ket{v_1} $, $ \ket{v_2} $ and on the second surface as $ \ket{v'_1} $, $ \ket{v'_2} $. On the left side of the figure the two edges are unbraided and on the right side they are wound once around each other.

Let us denote the Hilbert space associated with the $ i^{\text{th}} $ puncture as $ \mc{H}_i $. Then the state of the two punctures, either on $ S_1 $ or on $ S_2 $, resides in the tensor product Hilbert space: $ \mc{H}_1 \otimes \mc{H}_2 $. On the right side of the figure we see that braided the edges around each other has the same effect as exchanging the two punctures on the second surface. Now, in general, when we exchange two particles in any quantum system the total state of the system undergoes a unitary transformation. In three spatial dimensions particle exchange can only leads to a change in the phase of the wavefunction by $ +1 $ (for bosons) or $ -1 $ for fermions. However, in two spatial dimensions one can have more complex behavior (see for \eg \cite{Jain2007Composite}). The unitary operator associated with such an exchange need no longer be a trivial multiple of the identity.

\begin{figure}[h]
	\centering
	\includegraphics[height=80pt]{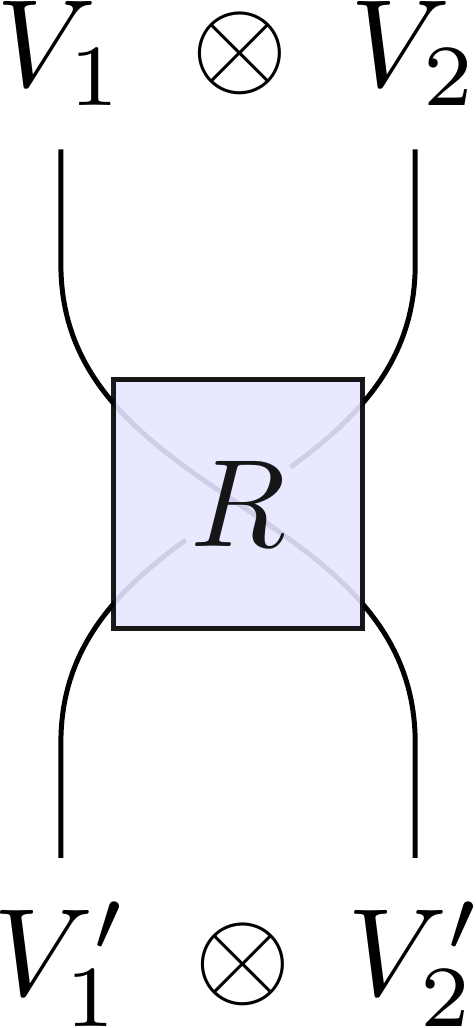}
	\caption{Effect of braiding two particles can be represented by the action of a unitary operator $ R $ on the two-particle Hilbert space}
	\label{fig:braiding-op}
\end{figure}

In general we would have the following relationship between the states of the two particles before and after exchange:
\begin{equation}\label{eqn:braiding}
	V_1' \otimes V_2' = R (V_1 \otimes V_2)
\end{equation}
where $ R $ is some unitary operator. Diagrammatically this is shown in the figure \autoref{fig:braiding-op}.

\begin{figure}[h]
	\centering
	\includegraphics[height=100pt]{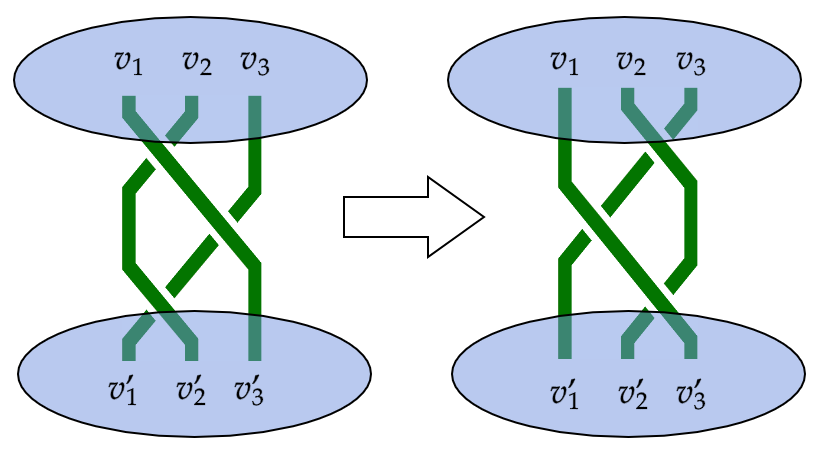}
	\caption{The two configurations shown here are connected to each other by a small diffeomorphism, known as a type IV Reidemeister move.}
	\label{fig:reidemeister-type-iv}
\end{figure}

The form of the unknown unitary $ R $ is, as yet, undetermined. We can now use the fact that an arbitary spin-network state should be invariant under small diffeomorphisms to determine an exact equation which must be satisfied by any $ R $. Consider the two configurations shown in \autoref{fig:reidemeister-type-iv}. Now instead of two punctures, we are considering the case where each surface contains three punctures. The two figures might seem different, but they are not. It is easy to convince oneself that one can slide the middle thread in such a way as to go from the configuration on the right to the one of the left, \emph{without} breaking or joining any threads. Such a transformation is nothing but a small diffeomorphism acting only on the middle thread!

However, while topologically the two configurations appear to be the same, the \emph{order} in which the pairwise unitary operator $ R $ acts on the three-particle state is different in both cases. On the left we see that the, in the first step, first two threads are braided while the first in untouched. This operation on the three particle Hilbert space can be denoted by the operator $ R_{12} \otimes \mb{1}_3 $, where the subscripts denote the particle on which the operation acts. The second step (on the left) can be represented as $ \mb{1}_1 \otimes R_{23} $ and the third one (on the left) is again $ R_{12} \otimes \mb{1}_3 $. The full unitary operation corresponding to the left hand side of \autoref{fig:reidemeister-type-iv} acting on the spin-network state can thus be expressed as:
\begin{equation}\label{eqn:yang-baxter-lhs}
	(R_{12} \otimes \mb{1}_3) (\mb{1}_1 \otimes R_{23}) (R_{12} \otimes \mb{1}_3) \ket{\Psi_\Gamma}
\end{equation}
In a similar way one can write the total unitary corresponding to the right hand side of \autoref{fig:reidemeister-type-iv}:
\begin{equation}\label{eqn:yang-baxter-rhs}
	(\mb{1}_1 \otimes R_{23}) (R_{12} \otimes \mb{1}_3) (\mb{1}_1 \otimes R_{23}) \ket{\Psi_\Gamma}
\end{equation}
While the two unitary operations shown in \eqref{eqn:yang-baxter-lhs} and \eqref{eqn:yang-baxter-rhs} \emph{appear} to be similar they are \emph{not} the same. However, if we insist that the state of quantum geometry represented by the given spin network should be invariant under small diffeomorphisms then the two unitary operations must have the same effect on the state $ \ket{\Psi_\Gamma} $. This equivalence can formally be written as an operator equation:
\begin{equation}\label{eqn:yang-baxter}
	(R_{12} \otimes \mb{1}_3) (\mb{1}_1 \otimes R_{23}) (R_{12} \otimes \mb{1}_3) = (\mb{1}_1 \otimes R_{23}) (R_{12} \otimes \mb{1}_3) (\mb{1}_1 \otimes R_{23})
\end{equation}
We see here that the particular spin-network state in question factors out because this equation should hold for \emph{all} spin-networks for the theory to be physically consistent, and we are finally left with just an operator equation. This equation is famous in the field of condensed matter and goes by the name of the Yang-Baxter equation \cite{Baxter2008Exactly} in honor of the two individuals most deeply associated with studying its solutions and recognizing its significance. It can be solved exactly to give us the exact form of the unitary operator $ R $. As we shall see in the next section the form of $ R $ is precisely that of a CNOT gate - a two qubit gate which is universal for quantum computation.

%Invariance of spin-network state under Reidemeister (IV) moves (statement of invariance w.r.t. (small) diffeomorphisms)
%
%\begin{equation}\label{eqn:braiding-relation}
%	\sigma_1 \sigma_2 \sigma_1 = \sigma_2 \sigma_1 \sigma_2
%\end{equation}

%\begin{equation}\label{eqn:braiding-relation-graphic}
%	\vcenter{\hbox{\includegraphics[scale=0.15]{ybe_1.pdf}}} =  \vcenter{\hbox{\includegraphics[scale=0.15]{ybe_2.pdf}}}
%\end{equation}

\section{Yang Baxter Equation and Quantum Computation}

It turns out to be the case that the manipulations of spin-network edges described in the previous section are precisely the same operations as those used to perform quantum computation using topological phases of two-dimensional condensed matter systems as the computing ``hardware''.

Let us assume for the time being that we can identify the Hilbert space of a puncture $ \mc{H} $ (c.f. \autoref{fig:braiding}) with of a spin $ 1/2 $ particle, \ie a qubit. In this case the operator $ R $ is a two qubit gate and can be expressed as a $ 4 \times 4 $ matrix. The exact solution for $ R $ in this case is given by the following expression \cite{Kauffman2004Braiding}:
\begin{equation}\label{eqn:cnot-gate}
\scalemath{1.0}{
	R =  \left(
	\begin{array}{cccc}
	1 & 0 & 0 & 0 \\
	0 & 1 & 0 & 0 \\
	0 & 0 & 0 & 1 \\
	0 & 0 & 1 & 0
	\end{array}
	\right)
}
\end{equation}
This operator acts of the state space of two qubits. The basis states of a single qubit can be written as $ \{\ket{0}, \ket{1}\} $. The basis states of a two qubit state can then be written as: $ \{ \ket{00}, \ket{01}, \ket{10}, \ket{11}\} $. The operator $ R $ given in \eqref{eqn:cnot-gate} acts on this state space. The effect of this operator can be described the circuit diagram shown in \autoref{fig:cnot-gate} which is repeated below for the reader's convenience.
\begin{figure*}[h]
	\centering
	\includegraphics[origin=c]{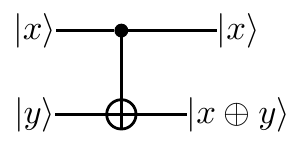}
	\caption*{Circuit diagram for a CNOT gate}
%	\label{fig:cnot-gate}
\end{figure*}

Now, it is a well known fact in classical computation that all of the gates used in Boolean logic - AND, OR, XOR, NAND and NOT - can be constructed given only the NAND gate. The NAND gate is thus said to be ``universal'' for classical computation. Similarly, using only a finite set of discrete two and one qubit gates one can construct \emph{any} unitary operator to arbitrary precision \cite[Ch. 4]{Nielsen2000Quantum}, by repeated application of the gates in the given set. One such set of universal gates consists of the CNOT gate, Hadamard gate, phase gate and the $ \pi/8 $ gate. Except for CNOT, the remaining gates all act on a single qubit. We list these gates below in the standard $ \{\ket{0}, \ket{1}\} $ basis:
\begin{subequations}\label{eqn:one-qubit-gates}
	\begin{align}
		\text{Hadamard}: & \quad H = \begin{pmatrix}
								1 & 1 \\ 
								1 & -1
								\end{pmatrix} \label{eqn:hadamard-gate} \\
		\text{Phase}: & \quad S = \begin{pmatrix}
								1 & 0 \\ 
								0 & i
								\end{pmatrix} \label{eqn:phase-gate} \\
		\pi/8: & \quad T = \begin{pmatrix}
								1 & 0 \\ 
								0 & e^{i\pi/4}
								\end{pmatrix} \label{eqn:piby8-gate}
	\end{align}
\end{subequations}
What we have discovered so far is that quantum diffeomorphism invariance of spin network states implies that these states naturally carry an implementation of the entangling two qubit CNOT gate. If, in addition, it would be possible to incorporate the gates mentioned in \eqref{eqn:one-qubit-gates} as some additional structure in spin-network states, then it would imply that spin-network states are universal for quantum computation. Already the existence of the CNOT gate hidden within the state space of loop quantum gravity, tells us that there is a deep connection between quantum computation and quantum gravity, confirming what we already know from various other independent approaches \cite{Almheiri2014Bulk,Mintun2015Bulk-Boundary,Pastawski2015Holographic,Freivogel2016Precursors,Harlow2017The-RyuTakayanagi} to the question of quantum gravity.

\section{Framed Edges and Single Qubit Operators}

It turns out that spin-networks can be endowed with additional topological structure in a natural manner and the degrees of freedom of this structure can then naturally be interpreted as the action of quantum gates acting on individual qubits. This structure turns out to be required due to two independent considerations. Let us explain these in turn.

In his famous paper \cite{Witten1989Quantum} Witten was the first to point the deep connections between (topological) quantum field theory and knot theory. In particular he showed the the expectation value of the Wilson loop functional evaluated on any knotted curve $ \gamma $ in $ SU(2) $ Yang-Mills theory in $ 2+1 $ dimensions is related to the knot invariant of $ \gamma $ known as the Jones polynomial. In this work he showed that a crucial step in the calculation involved the calculation of the self-linking number of a knot. This quantity is ill-defined if the knots are taken to be one-dimensional curves. In order to ``regularize'' the associated integral one has to introduce a \emph{framing} of the knot, \ie replacing the one-dimensional curve constituting the knot $ \gamma $ with a two-dimensional ribbon.

The second justification for using framed ribbons rather than one-dimensional strings comes from Smolin's seminal work \cite{Smolin1995Linking} linking topological quantum field theory with the field of loop quantum gravity which was in its infancy at the time. By considering the action of certain Wilson loop operators acting on punctures between spin-network edges and an arbitrary two-dimensional surface he was able to deduce that the punctures, and thereby the spin network edges, would have to be endowed with a framing.

Taken together, these two lines of argument are enough to convince oneself that the full state of loop quantum gravity should incorporate braided ribbon spin networks rather than the one-dimensional versions commonly encountered in the literature.

The careful reader would have noticed that we have used such framed ribbons, rather than one-dimensional lines, to represent spin network edges in several of the figures in this paper. This is not only for aesthetic convenience but also because ultimately we wish to work with framed or ribbon spin networks.

\begin{figure}[h]
	\centering
	\begin{subfigure}[t]{0.3\textwidth}
		\centering
		\includegraphics[height=60pt]{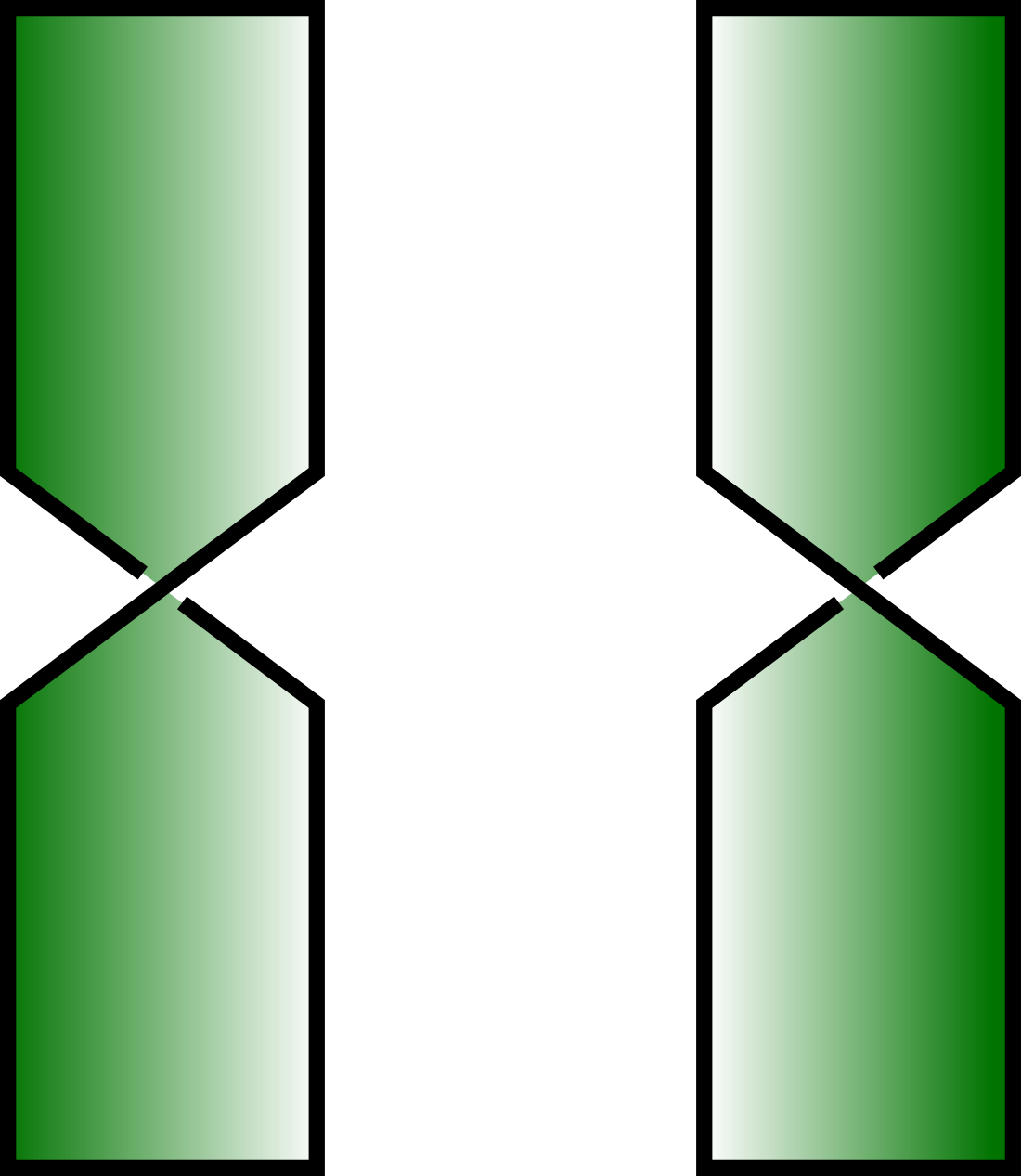}
		\caption{An example of a twist by $ \theta = + \pi $ and $ \theta = -\pi $}
		\label{fig:twists}
	\end{subfigure}
	\hfill
	\begin{subfigure}[t]{0.3\textwidth}
		\centering
		\includegraphics[height=60pt]{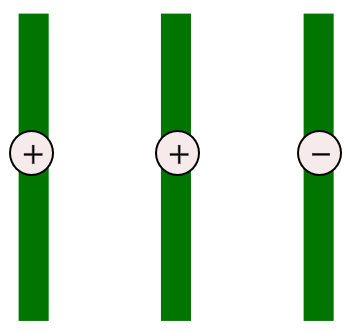}
		\caption{A configuration of three ribbons, with the first two having a twist $ \theta = +\pi $ and the last one with a twist $ \theta = -\pi $}
		\label{fig:twists-110}
	\end{subfigure}
	\hfill
	\begin{subfigure}[t]{0.3\textwidth}
		\centering
		\includegraphics[height=100pt]{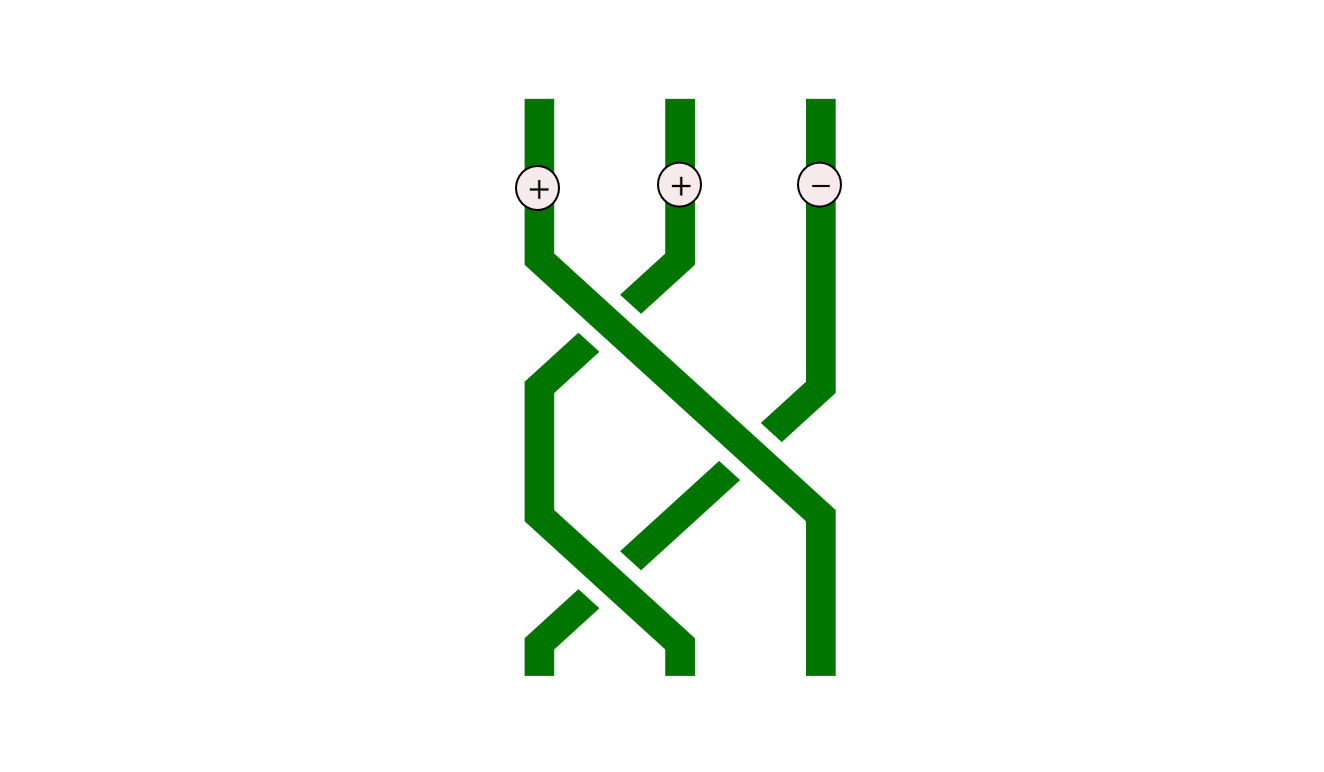}
		\caption{A configuration of three ribbons with braiding and twisting}
		\label{fig:braid-twists-110}
	\end{subfigure}
	\caption{Illustration of twisted braids}
\end{figure}

The moment we replace spin network edges with framed ribbons we realize that we have gained another topological degree of freedom in the form of \emph{twists} which the ribbon can have. Let us assume that the simplest possible twist can be if one end of the ribbon is rotated $ \theta $ degrees relative to the other edge in either a clockwise or counter-clockwise manner. $ \theta $ can, in principle, take any value between $ 0 $ and $ 2\pi $. Here, for simplicity, we will assume that $ \theta $ can only take values $ \pm \pi $. As we shall see later, this value also arises naturally when considering the application of discrete symmetries of spin-networks. \autoref{fig:twists-110} shows an example of a configuration with three ribbons, with the first two having a $ +\pi $ twist and third one with a $ -\pi $ twist. \autoref{fig:braid-twists-110} shows how the twisting and braiding operations can be combined.

It is natural that, as was the case with braiding of two edges, putting a twist in a single framed edge should also correspond to the action of some unitary operator on the given edge. It is not clear at this stage precisely which form this unitary operator should take out of the one-qubit gates listed in \eqref{eqn:one-qubit-gates}: the Hadamard, phase or $ \pi/8 $ gates, or perhaps some other one qubit unitary. What is clear is that in addition to CNOT, the twisting operation will give us at least one additional single qubit unitary. If somehow we were able to identify additional topological degrees of freedom which could be associated with the three single qubit gates listed in \eqref{eqn:one-qubit-gates} then we would be well placed to make the following proposition:

\begin{quote}
	\textbf{Proposition III}: \emph{The set of framed, braided spin network states provides a complete set of states required for universal quantum computation.}
\end{quote}
which could alternatively be stated:
\begin{quote}
	\textbf{Proposition III'}: \emph{The state space of loop quantum gravity is dense in the set of operators required for universal quantum computation.}
\end{quote}

We are now well placed to come to the central result of this work, which is the claim that the state space loop quantum gravity, extended by inclusion of braiding and twisting of framed edges, provides us with a source of generating GHZ states, which are a central resource for quantum error correcting codes. To do so we need only remind ourselves of the structure of the circuit which generates a GHZ state. This circuit is shown in \autoref{fig:GHZ-states}, which is reproduced below for the reader's convenience.
\begin{figure}[h]
	\centering
	\begin{subfigure}[t]{0.45\textwidth}
		\centering
		\includegraphics[scale=1.0]{qutrit_state}
		\caption{Quantum circuit for generating a qutrit state using CNOT gates}
		\label{fig:GHZ-states-2}
	\end{subfigure}
	\hfill
	\begin{subfigure}[t]{0.45\textwidth}
		\centering
		\includegraphics[height=50pt]{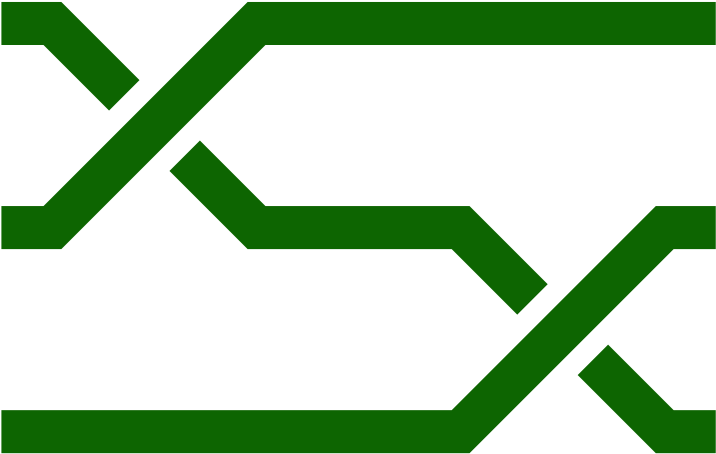}
		\caption{The braid operation corresponding to the quantum circuit for generating a GHZ state. Threads are numbered in increasing order from top to bottom. Time runs from left to right.}
		\label{fig:ghz-braid}
	\end{subfigure}
	\caption{Quantum circuit representation for the GHZ state (left) and the braid representation (right).}
\end{figure}
The interested reader can easily verify, using the explicit form of the $ R $ operator given in \eqref{eqn:cnot-gate} that the braid operation shown in \autoref{fig:ghz-braid} generates the GHZ state given that the first qubit starts in the state $ \ket{0} + \ket{1} $ and the second and third qubits start in the $ \ket{0} $ state. The shown braid configuration then corresponds to the unitary operator $ (R_{12} \otimes \mbb{1}_3)(\mbb{1}_1 \otimes R_{23}) $.

\section{Conclusion: Elementary Particles as Quantum Error Correcting Codes}

In this work we have made the following observations. First, in order to consider the full state space of loop quantum gravity, knotting and braiding of spin-network edges needs to be taken in account. Second, the action of the operator version of small diffeomorphisms is given by the Yang-Baxter equation whose solutions yield a two qubit unitary gate which is universal for quantum computation. And, finally, the inclusion of topological degrees of freedom allows us to generate states which are essential ingredients in quantum error correcting codes.

It is also interesting to note that the braid configurations shown in \autoref{fig:ghz-braid} are identical to those in the groundbreaking work by Sundance Bilson-Thompson \cite{Bilson-Thompson2005A-topological,Bilson-Thompson2006Quantum} wherein he identified these states with leptons belonging to the first generation of the Standard Model of elementary particles.
\begin{figure}[h]
	\centering
	\includegraphics[height=100pt]{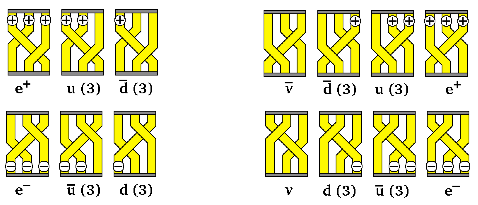}
	\caption{Assignments of configurations of three-ribbon twisted braids with various members of the first generation of leptons in the Standard Model (figure credit \cite{Bilson-Thompson2005A-topological,Bilson-Thompson2006Quantum})}
	\label{fig:bt-model}
\end{figure}
Considering \emph{all} possible configurations of the elements of the braid group with three strands with twists, one ends up with the set of configurations shown in \autoref{fig:bt-model}. By making the assumption that the $ \theta = \pm \pi $ twist on a ribbon corresponds to an electric charge of $ \pm 1/3 $ to that ribbon, we are find that there are precisely four configurations with total charge $ \pm 1 $, $ \pm 2/3 $ and $ \pm 1/3 $ each. There are two elements which are uncharged (without any twists on any of the ribbons). The charged elements can be identified with an electron (charge $ - 1 $), an up quark (charge $ +2/3 $), down quark (charge $ -1/3 $) and their antiparticles. After this assignment we are still left with two configurations for each value of the total charge. These two configurations are mirror reflections of each other. This motivates us to identify one set of these configurations with the left handed leptons $ (e^\pm_L, u_L, \bar u_L, d_L, \bar d_L) $ and the other with right handed leptons $ (e^\pm_R, u_R, \bar u_R, d_R, \bar d_R) $. The two uncharged configurations form a particle-antiparticle pair (applying the operation of braid concatenation to the two configurations gives us the identity braid) and are identified with the neutrino ($ \nu $) and the anti-neutrino ($ \bar v $).

There is a certain elegance in associating elementary particles with elements of the code words of a quantum error correcting code. The vacuum is filled with fluctuating quantum fields. However, we don't identify all of the possible fluctuations of the quantum vacuum with elementary particles. There is a certain element of stability that elementary particles possess. If we shoot an electron from one end of a long vacuum tube - from which the atmosphere has been evacuated - then, in the absence of any collisions with some stray particles passing by, we expect to find an electron arriving at the other end of the tube. This is true, even though the electron has to interact with the fluctuations of the quantum vacuum in its journey through the tube. In the language of error correction, such a state which is immune to fluctuations in its environment is known as a noiseless subsystem \cite{Zanardi1997Noiseless,Kribs2005Unified}. It is precisely these noiseless subsystems which serve as the elements of the \emph{code space}. It would only be natural if elementary particles were to be identified \cite{Kribs2005Geometry,Konopka2006Constrained} with the noiseless subsystems of any underlying quantum gravity theory.

It is important to note that several prior authors \cite{Bilson-Thompson2006Quantum,Wan2007Braid,Wan2009Effective,Vaid2010Embedding,Hackett2011aInvariants,Hackett2011bInvariants} have focused on the role of ribbon spin networks in loop quantum gravity. However, to the best of our knowledge this work (and another prior paper by the author \cite{Vaid2013Elementary}) are the only ones to have made the association between ribbon networks, elementary particles and quantum error correcting codes.

In recent work Freidel and colloborators \cite{Freidel2017Loop,Freidel2019Gravitational} have suggested assigning a $ U(1)^3 $ Kac-Moody algebra to the punctures where a spin network edge intersects a two-dimensional surface. They further argue that this implies that the edges of spin networks should be considered as tubes rather than as one dimensional curves. This sort of a structure was first suggested by the present author in a much older work \cite{Vaid2010Embedding} relying on arguments centered around discrete symmetries of spin networks. There might exist a direct link between the work of Freidel and co-workers and the present work. This is suggested by a close relation between solutions of the Yang Baxter equation and Kac-Moody algebras \cite{de-Vega1986Integrable}. This, however, must be left for future works. 

\bibliographystyle{jhep}
\bibliography{lqg-qec.bib}

%\printbibliography

%%\onecolumn\newpage
%
%\appendix
%%
%\section{Braiding from Discrete Symmetries}
%
%\begin{figure}[h]
%	\centering
%	\includegraphics[scale=0.20]{dihedral_group_ops}
%	\caption{Representation of the action of elements of the dihedral group ($D_2$) on a triangle. The triangle on the right hand side shows the axes of reflection. Thus $R_i$ stands for reflection across the $i^{\textrm{th}}$ axis of symmetry. The generators of rotations are $S_\theta$, with positive rotations being in the counterclockwise direction}\label{fig:DihedralGroup}
%\end{figure}
%
%\begin{figure}[h]
%	\centering
%	\includegraphics[scale=0.2]{punctured_sphere}
%	\caption{Invariance under discrete symmetries requires replacing 1D spin network edges with 2D tubes}
%\end{figure}
%
%\begin{figure}[h]
%	\centering
%	\includegraphics[scale=0.3]{triangle_map}
%	\caption{Map between triangle represented by an element of the braid group }\label{fig:TriangleMap}
%\end{figure}

\end{document}